# Tensor charge and anomalous magnetic moment correlation

Mustapha MEKHFI

International Center for Theoretical Physics Trieste, Italy

We propose a generalization of the upgraded Karl- Sehgal formula which relates baryon magnetic moments to the spin structure of constituent quarks, by adding anomalous magnetic moments of quarks. We first argue that relativistic nature of quarks inside baryons requires introduction of two kinds of magnetisms, one axial and the other tensoriel. The first one is associated with integrated quark helicity distributions $\Delta_i - \Delta_{\bar{i}}$ (standard) and the second with integrated transversity distributions $\delta_i - \delta_{\bar{i}}$. The weight of each contribution is controlled by the combination of two parameters, $x_i$ the ratio of the quark mass to the average kinetic energy and $a_i$ the quark anomalous magnetic moment. The quark anomalous magnetic moment is correlated to transversity and both are necessary ingredients in describing relativistic quarks. The proposed formula, then when confronted with baryon magnetic moments data with reasonable inputs, yields beside quark magnetic densities, anomalous magnetic moments enough large to not be ignored.

12.39.Ki, 13.40.Em, 24.10.Jv, 24.70.+s

I – Introduction.

Karl-Sehgal[1] formula (upgraded first by Cheng and Li[2] then by Di Qing[3] et al.) relating baryon magnetic moments to the spin structure of the constituent quarks takes into account the relativistic nature of quarks inside the parent nucleon. The upgraded formula by Di Qing et al. is a model independent, field theoretical relation which includes quark tensor charges in addition to the longitudinal spin part of the formula. At the relativistic level the transverse spin structure is an independent structure at , with respect to the longitudinal spin structure[4]. A straightforward but however lengthy way to obtain the formula is to expand quark field operators in nucleon matrix elements of quark currents in terms of a complete set of quark and antiquark wave functions. In performing such expansion, quark-antiquark pairs become operating if the baryon state is a Fock decomposition beyond the $q^3$ state. $|B\rangle = c_0 |q^3\rangle + \sum_{\alpha} c_{\alpha} |q^3 q\bar{q}\rangle_{\alpha} + \cdots$ Attempts have been made to generalize the formula by taking into account the contributions from quark-antiquark pairs in a constituent quark model with valence $q^3$ and sea $q^3 q\bar{q}$ mixing. It is found that pair creations only contribute a small amount to the magnetic moment of the proton ($-0.065$ $n.m$ with $n.m$ the nucleon magneton)[3]. It is to note that the inclusion of sea quarks by authors of reference 3 through the Fock space configuration is a tentative to include quark interactions into the scheme. In this paper we reconsider the problem of introducing interactions into the baryon magnetic moments formula by using a standard approach in which the baryon has the standard $q^3$ configuration. There are several possible sources of interactions which contribute to baryon magnetic moments. Exchange magnetic moments[5][6] ( they are generic in any interacting field theory), transition moments and individual anomalous magnetic moments (a.m.m) of quarks. Exchange magnetic moments contribute a non-additive piece to the baryon magnetic moments. This means that this contribution will add an additional term not proportional to the sum of individual quark magnetic moments. In the chiral quark model for instance, two- body exchange moments ( to consider only the leading )come from the exchange of one Nambu -Goldstone boson with one photon attached in all possible ways. A rough estimate of the size of exchange moments yields 0.010 $n.m$ [5].The exchange correction, being connected with exchange of charged pions, requires the presence of u and d quarks in the baryon and hence contributes only to the proton and the neutron. Transition moments add a (yet small) piece to the process $\Sigma \rightarrow \Lambda$ .Other contributions are due to anomalous magnetic moments of quarks. Such contributions on the contrary may be significant . Nonlinear chiral quark model for instance may be used to estimate the order of magnitude of the anomalous contribution. In fact one would expect an anomalous magnetic moment[7] of order $\frac{m_i^2}{\Lambda_{CSB}^2} \approx 10\%$ ( $m_i \approx 360 Mev$, $\Lambda_{CSB} \approx 1 Gev$ ) with $m_i$ being the constituent mass of the quark, supposed to be the effect of chiral symmetry breaking, and $\Lambda_{CSB}$ is the chiral symmetry breaking scale. There are several theoretical and experimental studies indicating quarks do have non negligible a.m.m. To fit the measured magnetic moment of the baryon octet, it is found that quarks must have a sizable a.m.m. In effect, non relativistic constituent quark model for light hadrons, with measured anomalous magnetic moments for the proton and the neutron respectively $a_p = 1.79$ and $a_n = -1.91$ yields the relations.

$$\frac{1}{3}(4\mu_u - \mu_d) = 2.79$$
$$\frac{1}{3}(4\mu_d - \mu_u) = -1.91$$
(1)

From which we infer measurable quantities $\frac{m}{1+a}$.

$$\frac{m_u}{1+a_u} = 338 \; Mev$$
$$\frac{m_d}{1+a_d} = 322 \; Mev$$
(2)

On the other hand, to fit hadron spectrum in the constituent quark model required masses of the up and down quarks to be of the order $m_u \simeq m_d = 420 \; Mev$. Such values of masses suggest a sizable anomalous magnetic moments of the order $a_u \simeq 0.24$, $a_d \simeq 0.30$ and a small difference $a_u - a_d \simeq 0.07$ to recover the isospin symmetry $m_u \simeq m_d$. Bicudo et al[8] have shown in several effective quark models, that in the case of massless-current quarks, chiral symmetry breaking usually triggers the generation of an anomalous magnetic for the quark of the order $a \simeq 0.28$. In the same spirit, Singh [9] has also proven that, in theories in which chiral symmetry breaks dynamically, quarks can have a large a.m.m. On the other hand, Köpp et al [10] have provided a stringent bound on the a.m.m from high-precision measurements at LEP, SLC, and HERA. In the second section we will give theoretical arguments showing that quark anomalous magnetic moments and tensor charges are necessarily correlated.

In the following we assume we have derived an effective lagrangian defined at the scale of low-energy magnetic moments after having integrated all unwanted fields. Constituent quarks have masses $m_i$ $i=u,d,s$ and do have anomalous magnetic moments from the term $\frac{a_i Q_i}{2m_i}\bar{\psi}\sigma_{\mu\nu}\psi F^{\mu\nu}$ in the effective lagrangian. Baryon magnetic moments $\vec{\mu}_N$ are composed of a contribution due solely to quark electric charges and their longitudinal spins (quark-antiquark pairs neglected) and other collective contributions such as exchange moments, transition moments and finally a contribution due the anomalous magnetic moments of quarks. The last two contributions are represented by dots.

$$\vec{\mu}_N = \langle PS | \sum_{i,\bar{i}} \frac{Q_i}{2} \int dr^3 \vec{r} \times \bar{\psi}_i \vec{\gamma} \psi_i | PS \rangle + \ldots \quad (3)$$

$Q_i, i=u,d,s$ are quark charges, $\psi_i(\psi_{\bar{i}})$ constituent quark (antiquark) fields and $|PS\rangle$ is the baryon ground state with momentum $P$ and spin polarization $S$. The spin structure of quarks is encoded in the axial and tensor charges, respectively denoted $\Delta i = \Delta_i + \Delta_{\bar{i}}$ and $\delta i = \delta_i - \delta_{\bar{i}}$ (the minus sign accounts for the odd charge conjugation parity of the transverse spin operator). The quark helicity density (antiquark) $\Delta_i(\Delta_{\bar{i}})$ is defined in the parton infinite momentum frame as $\Delta_i = \int dx \left[ q_{i\uparrow}(x) - q_{i\downarrow}(x) \right]$ with $q_{i\uparrow}(x), (q_{i\downarrow}(x))$, the probability of finding a quark with fraction $x$ of the baryon momentum and polarization parallel (anti parallel) to the baryon spin. It can also be shown to be related to the expectation value of the relativistic quark(antiquark) spin operator in the baryon

$$\langle PS | \int dx^3 \psi_i^\dagger \vec{\Sigma} \psi_i | PS \rangle = 2\Delta_i \vec{S} \qquad (4)$$

Similarly $\delta_i$ is given by the formula.

$$\langle PS | \int dx^3 \overline{\psi}_i \vec{\Sigma} \psi_i | PS \rangle = \vec{\delta}_i \qquad (5)$$

and can be shown to be related to the first moment of the quark transversity distribution $\delta_i = \int_0^1 dx [q_{\rightarrow i}(x) - q_{\leftarrow i}(x)]$ [11]. Similar expressions apply to the antiquark. Unpolarized quark distribution (well known), quark helicity distribution (known), and transversity distribution (unmeasured but calculated on lattice, and several other models), provide together, a complete description of the quark spin. To stress the difference between helicity and transversity, recall that if quarks moved non relativistically in the nucleon, $\delta_i(x)$ and $\Delta_i(x)$ would be identical as only large components of the fermion field are leading in which case $\overline{\psi} = \psi\dagger\gamma^0 \simeq \psi\dagger$ and both definitions (4) and (5) coincide. Another way of seeing this, is that rotations and Euclidean boosts (non relativistic case) commute and a series of boosts and rotations can convert a longitudinally polarized nucleon into a transversely polarized nucleon at infinite momentum. So the difference between transversity and helicity distributions reflects the relativistic motion of quarks inside the nucleon.

To express baryon magnetic moments in terms of spin degrees of freedom we compute (3) using the field current $\vec{j}_i = \overline{\psi}_i \vec{\gamma} \psi_i$ and assume the ground state of the baryon to have a vanishing non-relativistic orbital magnetic moment. To this end it is useful to decompose the quark current into two distinct pieces using Gordon decomposition and to not expand quark field operators in terms of a complete set of quark and antiquark wave functions as in the previous cited work . The convection current part and the spin current part contribute differently, giving respectively.

$$\frac{x_i \mu_i}{2(1+x_i)}(\Delta_i - \frac{\delta_i}{x_i})$$
$$\frac{x_i \mu_i}{2}(\Delta_i + \frac{\delta_i}{x_i})$$
(6)

where $x_i = \frac{m_i}{\langle E_i \rangle}$ is the ratio of the constituent quark mass to the average kinetic energy of the quark in the baryon ground state. Adding antiquarks and denoting $\vec{\mu}_N = \langle P \uparrow | \vec{\mu}_N | P \uparrow \rangle$ we get.

$$\mu_N = \sum_{i=u,d,s} \mu_i W_i + ...,$$
$$\mu_i = \frac{Q_i}{2m_i},$$
$$2\frac{W_i}{x_i} = \frac{1}{(1+x_i)}(\Delta_i - \Delta_{\bar{i}} - \frac{\delta i}{x_i}) + (\Delta_i - \Delta_{\bar{i}} + \frac{\delta i}{x_i})$$
(7)

.

Equation (7) is the upgraded Karl-Sehgal formula cited in reference 3 but obtained in another rearrangement of terms. Equation (7) is the weighted sum of two distinct combinations $(\Delta_i - \Delta_{\bar{i}} - \frac{\delta i}{x_i})$ and $(\Delta_i - \Delta_{\bar{i}} + \frac{\delta i}{x_i})$. The former combination shrinks to zero in the non relativistic limit. The latter combination survives the non relativistic limit and has the advantage that it is the only one which will be affected by the anomalous magnetic moments of quarks. Let us cite by the way a misuse of the Gordon decomposition which occurred twice in literature [12] and [13]. In Gordon decomposing the magnetic moment, the spin part takes the form $\vec{\mu}\,|_{spin} \propto \int \frac{1}{2m}\vec{r}\times\partial_v(\overline{\psi}\vec{\sigma}^v\psi)$ where $\vec{\sigma}^v$ is a vector which components are $\sigma^{iv}$. The spatial derivative $\partial_i$ gives (after neglecting a total derivative) the term $\int \frac{1}{2m}\overline{\psi}\vec{\Sigma}\psi$ [a] while the time derivative $\partial_0$ gives a non vanishing contribution, as quark fields do depend on time. What induced the above authors in error is probably the fact that the quark field $\psi(\vec{x},t)$ being interacting with gluons can not be expanded in terms of Dirac spinors in a free manner but can still be expanded at a given time say t=0, hence $\psi(\vec{x},0)$. We can proceed this way but after having performed the time differentiation if Gordon decomposition is to be used. In the appendix we give a correct computation of the spin part.

II – Tensor charge and anomalous magnetic moment correlation. Let us have a close look to formula (7). This formula has an insufficiency. It leads to an absence of magnetism in the ultra-relativistic limit due in part to the that, it is the average energy of the quark inside the baryon that builds up the intrinsic magnetic moment and not the constituent mass $m_i$ i.e. $x\mu_i \sim \frac{m_i}{\langle E_{i0}\rangle}\frac{1}{m_i} = \frac{1}{\langle E_{i0}\rangle}$ which goes to zero for infinite kinetic energy. The reduction factor $x$ is explicit in (7) and is simply the Lorentz-Fitzgerald contraction length due to the relativistic boost as the magnetic moment is a vector (space components of a four vector). On the other hand, tensor charges in the formula, being there to account for constituent quark masses (the mass term $m\overline{\psi}\psi$ flips helicity and hence involves transversity), should also disappear in this limit. We have indeed $\mu_N |_{ultra} \sim (-\delta i + \delta i) = 0$. The absence of magnetism in this limit suggests that formula (7) does have a missing term and that this term is associated with the anomalous magnetic moment of the quark. Why did we say that the anomalous magnetic moment of the quark is the missing term?. Formula (7) is a relativistic formula which describes how a magnetic photon couples to quarks being spinning point like objects. It also says that this coupling is decreasing with energy due to the reduction factor . On the other hand we know from quantum mechanics that particles of definite energy and momentum are not localized. It then follows a possible current in the lagrangian of the form[b]

$$\frac{\partial_\alpha \bar\psi \sigma^{\alpha\beta} \psi_\beta}{m} \qquad (8)$$

---

[a] The error made by cited authors is that they only retain the $\vec{\Sigma}$ term

Perturbatively, for a photon to probe such a current, a quark should radiate a field ( gluon or goldstone boson or whatever ) at position $x$ and reabsorbed at a distant position $y$, once it interacts with the photon ( vertex interaction and not a self-energy interaction ).In this process the probing photon sees the quark as an extended object or rather an electric current circulating in the area of the extension .This is what we call "anomalous" magnetism. The correlation of the anomalous magnetic moment to the tensor charge is suggested by the structure of the current (8) which, as the mass term, flips helicity.

---

[b] Differentiation of the field is non zero only if the field has a spatial and/or temporal extension. Point like objects have a current without derivatives such as $\bar\psi \gamma_\alpha \psi$ for instance.

Adding quark anomalous magnetic moments of quarks to formula (7), this one generalizes to (see appendix).

$$\mu_N = \sum_{i=u,d,s} \mu_i W_i + ...,$$

$$2\frac{W_i}{x_i} = \frac{1}{(1+x_i)}(\Delta_i - \Delta_{\bar{i}} - \frac{\delta i}{x_i}) + (1+a_i)(\Delta_i - \Delta_{\bar{i}} + \frac{\delta i}{x_i})$$

(9)

There is another different way of seeing that quark anomalous moments are missing. Let us rearrange formula (7) as this.

$$2W_i = A_i(\Delta_i - \Delta_{\bar{i}}) + B_i(\delta_i - \delta_{\bar{i}})$$

(10)

Parameters $A_i$ and $B_i$ are expressed in terms of $x_i$.

$$\frac{A_i}{x_i} = \frac{1}{1+x_i} + 1$$

$$B_i = -\frac{1}{1+x_i} + 1$$

(11)

Being functions of only one common parameter $x_i$, $A_i$ and $B_i$ are not independent parameters. Hence, these parameters could not distinguish between the contribution to baryon magnetic moments coming from helicities and the contribution coming from transversities, while these are supposed to be independent contributions in a relativistic regime. In general one may imagine that having two different spin structures in relativistic physics, namely, the longitudinal spin $\Delta_i, \Delta_{\bar{i}}$ and the transverse spin $\delta_i, \delta_{\bar{i}}$, quarks necessarily would carry two different magnetisms respectively of the form

$\mu_i A_i (\Delta_i - \Delta_{\bar{i}})$ and $\mu_i B_i (\delta_i - \delta_{\bar{i}})$ [c]. So in the relativistic case, the most general contribution to the baryon magnetic moments of quarks and antiquarks would be of the form(10) but where $A_i$ and $B_i$ are two independent parameters. Identifying coefficients of axial and tensoriel magnetic densities in both (9) and (10) we get two independent parameters.

$$\frac{A_i}{x_i} = \frac{1}{1+x_i} + 1 + a_i$$
$$B_i = -\frac{1}{1+x_i} + 1 + a_i$$
(12)

We understand that the introduction of anomalous magnetic moment is a necessary requirement of relativity, otherwise parameters $A_i$ and $B_i$ would be dependant parameters ( i.e. depend only on one parameter $x_i$ ) which means that helicity and transversity would no longer be two different spin structures in relativity. On the other it becomes also clear in this approach, that the quark anomalous magnetic moment is correlated to the quark transversity. Such a correlation is manifest at the ultra relativistic at which $W_i$ function in (9) takes the form.

$$2W_i = a_i (\delta i)_{ultra}$$
(13)

where $(\delta i)_{ultra} = \frac{2}{3}\delta_i^{NR} = \frac{2}{3}\Delta_i^{NR}$ is the ultra relativistic limit according to the solution of equations(18). This limit makes it explicit that quark anomalous magnetic moments together with tensor charges dominate the ultra relativistic regime.

---

[c] Hereafter we will call the first, axial magnetism ( although it is not the axial charge $\Delta_i + \Delta_{\bar{i}}$ (sum) which is involved but $\Delta_i - \Delta_{\bar{i}}$ (difference)) and the second, tensoriel magnetism.

III – Baryon magnetic moments analysis.

To include all baryons in the proton scheme, we assume SU(3) flavor symmetry. This enables us to write all baryon magnetic moments in terms of $W_i$ associated to the proton.

$$\mu(p) = \mu_u W_u + \mu_d W_d + \mu_s W_s$$
$$\mu(n) = \mu_u W_d + \mu_d W_u + \mu_s W_s$$
$$\mu(\Lambda) = \frac{\mu_u + \mu_d}{6}(W_u + 4W_d + W_s) + \frac{\mu_s}{3}(2W_u - W_d + 2W_s)$$
$$\mu(\Sigma^+) = \mu_u W_u + \mu_d W_s + \mu_s W_d$$
$$\mu(\Sigma^0) = \frac{\mu_u + \mu_d}{2}(W_u + W_s) + \mu_s W_d \tag{14}$$
$$\mu(\Sigma^-) = \mu_u W_s + \mu_d W_u + \mu_s W_d$$
$$\mu(\Xi^0) = \mu_u W_d + \mu_d W_s + \mu_s W_u$$
$$\mu(\Xi^-) = \mu_u W_s + \mu_d W_d + \mu_s W_u$$
$$\mu(\Sigma^0 \to \Lambda^0) = -\frac{(\mu_u - \mu_d)(W_u - 2W_d + W_s)}{2\sqrt{3}}$$

A consequence of the SU(3) symmetry is that magnetic moments can be written with 4 parameters, instead of the 6 parameters $\mu_i$ and $W_i$. Denoting the four parameters $c_0, c_3, c_8$ and $r$ we get

$$\mu(p) = c_0 + 2c_8 + 2c_3$$
$$\mu(n) = c_0 + 2c_8 - 2c_3$$
$$\mu(\Lambda) = c_0 - (3r-1)c_8$$
$$\mu(\Sigma^+) = c_0 + (3r-1)c_8 + (1+\frac{1}{r})c_3$$
$$\mu(\Sigma^0) = c_0 + (3r-1)c_8$$
$$\mu(\Sigma^-) = c_0 + (3r-1)c_8 - (1+\frac{1}{r})c_3$$
$$\mu(\Xi^0) = c_0 - (3r+1)c_8 - (1-\frac{1}{r})c_3$$
$$\mu(\Xi^-) = c_0 - (3r+1)c_8 + (1-\frac{1}{r})c_3$$
$$\mu(\Sigma^0 \to \Lambda^0) = -\frac{(3-\frac{1}{r})c_3}{\sqrt{3}} \tag{15}$$

where

$$\begin{aligned}
c_0 &= (\mu_u + \mu_d + \mu_s)(W_u + W_d + W_s)/3 \\
c_3 &= (\mu_u - \mu_d)(W_u - W_d)/4 \\
c_8 &= (\mu_u + \mu_d - 2\mu_s)(W_u + W_d - 2W_s)/12 \\
r &= \frac{W_u - W_d}{W_u + W_d - 2W_s}
\end{aligned} \tag{16}$$

We have a system of four equations (once coefficients $c^s, r$ are fixed) but six independent variables. To solve it we need two assumptions. To this end we first rewrite the system in terms of only five (new) variables $\frac{\mu_u}{\mu_d}, \frac{\mu_u}{\mu_s}, \mu_i W_i = \tilde{W}_i$ and then making the standard assumption $\frac{\mu_u}{\mu_d} = -2$, we end up with a soluble system ( four equations and four variables). Putting $f = \frac{3rc_0}{c_3 - 3rc_8}$ we get.

$$\begin{aligned}
\tilde{W}_u &= \frac{4}{9}\frac{c_3}{r}(f+1+3r) \\
\tilde{W}_d &= -\frac{2}{9}\frac{c_3}{r}(f+1-3r) \\
\tilde{W}_s &= -\frac{1}{9}\frac{9rc_8 - c_3}{r}(f-2) \\
\frac{\mu_s}{\mu_d} &= \frac{c_3 + 3rc_8}{2c_3} - 2\frac{c_3 - 3rc_8}{2c_3}
\end{aligned} \qquad (17)$$

It remains to fix one of the parameters, say $\mu_u$ to get access to the $W_i^s$ from the experimental data $c_i^s, r$. Parameters $x_i$ which appear in equation (18) remain undetermined. To fix them we call for Melosh-Wigner rotation reductions of nucleon spin which are due to quarks being relativistic particles inside baryons[d]. Indeed we have following relations between Pauli and Dirac spinors

$$\begin{aligned}
\bar{u}_{s'}(k)\gamma^3\gamma^5 u_s(k) &= M_A \; \chi_{s'}^\dagger \sigma_3 \chi_s \\
\bar{u}_{s'}(k)\gamma^0\gamma^3\gamma^5 u_s(k) &= M_T \; \chi_{s'}^\dagger \sigma_3 \chi_s
\end{aligned}$$

$M_A, M_T$ being the known Melosh[14][15][16] rotations. These rotations are shown to verify identities which in terms of spin densities take the form .

---

[d] We have already invoked reduction of magnetic moments as consequence of Lorentz boost. Here it is rather, the reduction of the spin which matters as we know the value of the spin before and after the reduction hence the value of the reduction factor $x_i$.

$$(1+x_i)\Delta_i^{NR} = \Delta_i + \delta_i$$
$$\Delta_i^{NR} + \Delta_i = 2\delta_i$$
(18)

The second expression is obtained if one assumes, in addition, that quark momentum distributions of nucleon ground state is spherically symmetric, that is $\langle k^2_\perp \rangle = 2\langle k^2_3 \rangle$. These relations serve to extract parameters $x_i$ from knowledge of naïve quark model spin densities $\Delta_i^{NR}$ and relativistic spin densities $\Delta_i$.

IV – Numerical applications.

As far, we have the experimental data [17] for seven magnetic moments ( $\mu(\Sigma^0)$ is not available) and one transition moment $\mu(\Sigma^0 \rightarrow \Lambda^0)$. Various calculations estimated collective contributions to be small. They are however shown to be necessary in order to satisfy sum rules which are consequences of SU(3) symmetry. In doing so one gets a best fit to the baryon magnetic moments and avoids introduction of artificial errors as in the Karl analysis of Karl-Sehgal equations. These corrections to magnetic moments however concern only the proton, the neutron and the transition $\Sigma^0 \rightarrow \Lambda^0$ and are accounted for by adding a constant to their magnetic dipole moments.

$$\mu(p) = ... + V$$
$$\mu(n) = ... - V$$
$$\mu(\Sigma^0 \rightarrow \Lambda^0) = ... - \frac{1}{\sqrt{3}}V$$

The following numeric values[18]

$$c_0 = 0.054 \ n.m$$
$$c_3 = 1.046 \ n.m$$
$$c_8 = 0.193 \ nm$$
$$r = 1.395 \ n.m$$

correspond to a best fit with $\chi^2/d.o.f = 1.3$. We will base our numeric analysis of baryon magnetic moments on these values. Parameter $V = 0.266 \pm 0.01 \ n.m$ serve to predict $\mu(\Sigma^0)$ but is of no relevance to $W_i^s$ as it describes only collective effects. Inserting these values into (17) we get.

$$\tilde{W}_u = 2.04$$
$$\tilde{W}_d = 0.37$$
$$\tilde{W}_s = 0.12$$
$$\frac{\mu_s}{\mu_d} = 0.66$$

To estimate $x_i$ we use equation (18) and write ($x_u = x_d = x$)

$$x = \frac{9}{10}(\Delta_u - \Delta_d) - \frac{1}{2} = \frac{9}{10}(g_A - \eta) - \frac{1}{2} = 0.60 \qquad (19)$$

Where values $\Delta_u^{NR} = 4/3$, $\Delta_d^{NR} = -1/3$ have been used together with the Bjorken sum rule $g_A = \Delta u - \Delta d = 1.27$ and the result from HERMES collaboration [19] $\eta = \Delta \bar{u} - \Delta \bar{d} = 0.05$. Remember that in changing variables from $W_i^s$ to $\tilde{W}_i^s = \mu_i W_i^s$ we reduced the number of variables by one unit and were able to solve the system of equations. The price we paid is the unknown parameter $\mu_u$ (other $\mu_i$ are linked to this one) still present in our formula.

$$2\frac{\tilde{W}_i(B)}{\mu_i x_i} = \frac{1}{(1+x_i)}(\Delta_i - \Delta_{\bar{i}} - \frac{\delta i}{x_i}) + (1+a_i)(\Delta_i - \Delta_{\bar{i}} + \frac{\delta i}{x_i}) \qquad (20)$$

One look at the above formula as a family of parametric sheets (magnetic sheets) in the 3-dimensional space of coordinates $\Delta_i - \Delta_{\bar{i}}, \delta i, a_i$, the parameter being $\mu_i$. To continue we select one magnetic sheet which corresponds to $\mu_u = 2.38\,n.m$ ($m_u \simeq 263\,Mev$). This value is obtained from the non relativistic form of (22) with vanishing anomalous magnetic moments[e 18].

$$\mu_u = \frac{2\tilde{W}_d + \tilde{W}_u}{g_A - 2\eta} = 2.38\,n.m \qquad (21)$$

To investigate equation (20) further, we make the approximation $a_u \simeq a_d = a$ .( this is suggested by the estimation we did in (2) ).Subtracting u quark contribution from d quark contribution in (20) and plug in the experimentally measured quantity $2\eta - g_A = (\Delta_d - \Delta_{\bar{d}}) - (\Delta_u - \Delta_{\bar{u}})$ we get.

$$\frac{2}{x\mu_u}(2\tilde{W}_d + \tilde{W}_u) = \frac{1}{(1+x)}(g_A - 2\eta + \frac{\delta d - \delta u}{x}) + (1+a)(g_A - 2\eta - \frac{\delta d - \delta u}{x}) \qquad (22)$$

All parameters in the above equation are fixed except the values of tensor charges, and the average over the up and the down quark anomalous magnetic dipole moment $a$.

---

[e] The constituent quark mass $m_i$ ( or equivalently $\mu_i$) is a free parameter in our approach .It is not fixed by baryon magnetic moments data. The best way to estimate its order of magnitude is to take the non relativistic limit of our formulas.

Tensor charges are not measured yet, but are estimated in various models. Using tensor charges from various models[f] such as, Chiral soliton (NJL)[20] [21],[22] Lattice(LAT)[23], Melosh-Wigner (M-W)[24], Valence sea quark mixing model (VSQMM) [1], Quark soliton model (QS)[25], Bag model (BAG)[26] [27], the constituent quark model with Goldstone boson effects (CQ)[28], Qcd sum rules(SR)[27], we compute anomalous magnetic moments $a = a_u \simeq a_d$ of quarks, averaging over models, and also axial magnetic densities. Our results are as follows.

$$a_u \simeq a_d \simeq 0.38$$
$$\Delta_u - \Delta_{\bar{u}} \simeq 0.78 \quad (23)$$
$$\Delta_d - \Delta_{\bar{d}} \simeq -0.39$$

We may use above axial magnetic densities $\Delta_u + \Delta_{\bar{u}} = 0.83$ ; $\Delta_d + \Delta_{\bar{d}} = -0.44$ to infer sea quark polarizations

$$\Delta_{\bar{u}} \simeq 0.03$$
$$\Delta_{\bar{d}} \simeq -0.05 \quad (24)$$

to compare with those obtained in the latest publication which used a standard non relativistic zero anomalous magnetic moment approach[18] $\Delta_{\bar{u}} \simeq -0.01$, $\Delta_{\bar{d}} \simeq -0.06$.

Predictions for the strange quark suffer from lack of experimental information. To get an order of magnitude of the anomalous magnetic moment of the strange quark we consider following hypothesises, usually applied not only to strange quarks but to all flavours simultaneously[29]

---

[f] If tensor charges for quarks get measured accurately, then one may use formula (22) to extract precise values of quark anomalous magnetic moments and vice versa.

Hypothesis A) Strange antiquarks in a polarized baryon are generated entirely by the perturbative splitting of gluons $g \to s\bar{s}$. In such a case, it is reasonable to expect $\Delta_{\bar{s}} \approx \Delta_s$ that is a vanishing axial magnetism, $\Delta_s - \Delta_{\bar{s}} = 0$. In this case $\tilde{W}_s$ of (20) takes the simpler form.

$$2\frac{\tilde{W}_s(B)}{\mu_s} = (1 + a_s - \frac{1}{(1+x_s)})\delta s \qquad (25)$$

Hypothesis B) Strange antiquarks in a polarized baryon reside entirely in a cloud of spin –zero strange mesons. In this case, strange antiquarks have no net polarization, i.e., $\Delta_{\bar{s}} = 0$, so that $\Delta_s - \Delta_{\bar{s}} = \Delta_s + \Delta_{\bar{s}}$. Equation (20) become in this case

$$2\frac{\tilde{W}_i(B)}{\mu_i x_i} = \frac{1}{(1+x_i)}(\Delta_i + \Delta_{\bar{i}} - \frac{\delta i}{x_i}) + (1+a_i)(\Delta_i + \Delta_{\bar{i}} + \frac{\delta i}{x_i}) \qquad (26)$$

We cannot extract $x_s$ from Melosh-Wigner rotation as we did for the light quarks simply because there is no transformation relation between $\Delta_{\bar{s}}(\delta_{\bar{s}})$ and $\Delta_s^{NR}(\delta_s^{NR})$). But the strange quark being heavier that the up and the down quark, gets less kinetic energy, so we may take for illustration $x_s \simeq 0.8 - 1$. Our results for the strange quark are displayed in Tableau 1.

| $\mu_s = 0.33\mu_u = 0.86$ | $\delta s$ | $x_s \simeq 0.8$ $a_s$ | $x_s \simeq 1$ $a_s$ |
|---|---|---|---|
| QS | -0.01 | 1.41 | 0.9 |
| VSQMM | -0.024 | 1.17 | 0.76 |
| LATT | -0.046 | 0.91 | 0.58 |
| Chiral Quark potential[30] | -0.133 | 0.38 | 0.20 |

**Tableau 1: Strange quark anomalous magnetic moments in model B for ratios $x_s \simeq 0.8 - 1$ and for various models.**

We do not display results in model A, because they gave unrealistic values for the strange quark anomalous magnetic moments. On the other hand we note that only the chiral quark potential model seems to give acceptable values $a_s \simeq 0.20 - 0.38$ while the lattice model producing a too small value for the tensor charge, yields a too high unacceptable anomalous magnetic moment $a_s \simeq 0.91$. It leads however to moderately reasonable value $a_s \simeq 0.58$ for strange quarks nearly at rest $x_s \simeq 1$.

V – Conclusion

Magnetic moments of the nucleon are static properties ( nucleon at rest). The quark inside the nucleon are nevertheless strongly bound relativistic objects. Being relativistic, the spin structure of quarks involves in general, both quark helicity distributions and quark transversity distributions. Transversity distributions encode relativistic effects of quarks inside the nucleon. We have shown in this study that since relativity requires existence of two independent spin structure, one longitudinal and the other transverse, it then follows, the existence of two independent magnetisms which we may call respectively axial and tensoriel. The contribution of each component is weighted by two independent parameters namely $0 \prec x_i \prec 1$ the ratio of the quark constituent mass to the quark average kinetic energy, and the anomalous magnetic moment $a_i$. Hence the quark anomalous magnetic moment $a_i$ is strongly correlated to the tensor charge $\delta i$ and this correlation is made more explicit in the ultra relativistic limit. Upgraded Sehgal-Karl-Chen formula relating baryon magnetic moments to the quark spin is a relativistic formula which necessarily includes quark tensor charges, but according to above considerations such formula is lacking essential ingredients which are quark anomalous magnetic moments which are correlated to tensor charges. To get a consistent formula for baryon magnetic moments we do add the missing part. We then confront our formula with baryon magnetic moments data using reasonable inputs such as $\frac{\mu_u}{\mu_d} = -2$, non relativistic limit to extract the quark mass $m_u \simeq 263\ Mev$, Melosh – Wigner rotation reductions of nucleon spin to estimate $x_u$, $x_d$ and tensor charges from various model computations. Anomalous magnetic moments of the u, d, and s quarks are evaluated, $a_u \simeq a_d \simeq 0.38$, $a_s \simeq 0.20 - 0.38$ and turn out to be enough large to not be ignored in any reliable analysis. Axial magnetic densities $\Delta_i - \Delta_{\bar{i}}$ for the up and down quarks or equivalently sea antiquark polarizations are also extracted and are different from values obtained in standard analysis of baryon magnetic moments. Our values are

$\Delta_{\bar{u}} \simeq 0.03$, $\Delta_{\bar{d}} \simeq -0.05$ to be compared with values obtained in an approach without quark anomalous magnetic moments nor quark tensor charges $\Delta_{\bar{u}} \simeq -0.01$, $\Delta_{\bar{d}} \simeq -0.06$.

**Appendix**

To prove the anomalous part of formula (9), we Fourier transform the anomalous part of the magnetic moment operator. We consider only one flavor and no antiquark to simplify notations

$$\vec{\mu}_N \big|_{anomalous} = -\frac{aQ}{2}\left\langle P\uparrow \bigg| \int \frac{\partial}{\partial \vec{q}} \times (\bar{\psi}(k')\frac{\vec{\sigma}^\nu}{2m}\psi(k)q_\nu)\frac{d^3p}{(2\pi)^3}\bigg|_{q=0} \bigg| P\uparrow \right\rangle$$

with $\vec{q} = \vec{k} - \vec{k}'$, $\vec{p} = \dfrac{\vec{k}+\vec{k}'}{2}$ and $\vec{\sigma}^\nu$ is a vector whose components are $\sigma^{i\nu}$. Then write

$$-\sigma^{i\nu} q_\nu = -(\vec{q}\times\vec{\Sigma})^i + i\vec{\alpha}q_0$$
$$\sigma^{ij} = \epsilon^{ijk}\Sigma_k$$
$$\sigma^{io} = -i\alpha^i$$

Differentiate each term of the above expression

$$i\frac{\partial}{\partial \vec{q}}\times(\bar{\psi}\vec{\alpha}q_0\psi)\bigg|_{q=0} = i\frac{\vec{k}\times\bar{\psi}\vec{\alpha}\psi}{k^0}$$

$$= \frac{m}{k_0}\bar{\psi}\vec{\gamma}\gamma_5\psi - \bar{\psi}\vec{\gamma}\gamma_5\gamma_0\psi$$

$$= \frac{m}{k_0}\psi^\dagger\vec{\Sigma}\psi - \bar{\psi}\vec{\Sigma}\psi$$

$$-\frac{\partial}{\partial \vec{q}}\times(\vec{q}\times\bar{\psi}\vec{\Sigma}\psi)\bigg|_{q=0} = 2\bar{\psi}\vec{\Sigma}\psi$$

To get the second term in the first equation we used the identity

$$\frac{\vec{\gamma}(\vec{\gamma}.\vec{k}) + (\vec{\gamma}.\vec{k})\vec{\gamma}}{2} = i(\vec{\gamma}\times\vec{k})\gamma_0\gamma_5$$

Using the definition of the tensor and axial currents

$$\langle PS | \int dx^3 \psi_i^\dagger \frac{\vec{\Sigma}}{2} \psi_i | PS \rangle = \Delta_i \vec{S}$$

$$\langle PS | \int dx^3 \bar{\psi}_i \vec{\Sigma} \psi_i | PS \rangle = \vec{\delta}_i$$

we get for the anomalous part

$$\mu_N \big|_{anomalous} = \frac{aQ}{4m} \langle P \uparrow | \int \frac{m}{k_0} \psi^\dagger \vec{\Sigma} \psi + \bar{\psi} \vec{\Sigma} \psi ) \frac{d^3k}{(2\pi)^3} | P \uparrow \rangle$$

$$= aQ \langle P \uparrow | \frac{\vec{S}}{2k_0} + \frac{1}{2m}(\frac{1}{2}\vec{\delta}) | P \uparrow \rangle$$

$$= \frac{axQ}{4m}(\Delta + \frac{\delta}{x})$$

$$= \frac{ax\mu}{2}(\Delta + \frac{\delta}{x})$$

References

[1] L.M. Sehgal, Phys.Rev.D **10**, 1663 (1974); G. Karl, Phys.Rev.D **45**, 247 (1992)

[2] T.P. Cheng and L.F. Li, Phys.Lett.B **366**, 365 (1996).

[3] Di Qing, Xiang-Song Chen, and Fan Wang, Phys.Rev.D **58**, 114032 (1998).

[4] X. Artru and M. Mekhfi, Z.Phys.C **45**, 669 (1990).

[5] Lee Brekke, Annals of physics **240**, 400 (1995).

[6] J. Franklin, Phys. Rev.D **66**, 033010 (2002); Phys.Rev.D **30**, 1542 (1984).

[7] A. Manohar and H. Georgi, Nucl.Phys.B **234**, 189 (1984).

[8] Pedro J.de A. Bicudo, J. Emilio F.T. Ribeiro, and Rui Fernandes, Phys.Rev.C **59**, 1107

[9] J.P. Singh, Phys.Rev.D **31**, 1097 (1985).

[10] G. Köpp, D. Schaile, M. Spira, and P.M. Zerwas, Z. Phys.C **65** (1995)

[11] R.L. Jaffe and X. Ji, Phys.Rev.Lett **67**, 552 (1990).

[12] X. Artru, Proceedings of RIKEN BNL Research center Worshop on future Measurements, Vol.29, (2000), edited by D.Boer and M.G.Perdekamp hep-ph/0207309

[13] X-S Chen, Di Qing, W-M Sun, H-S Zong, and F. Wang, Phys.Rev.C **69**, 045201 (2004).

[14] H.J. Melosh, Phys.Rev.D **9**, 1095 (1974).

[15] F.E. Close, in *An introduction to quarks and partons*, **1979**, edited by Academy Press, London, Chap. 6 and 11

[16] I.Schmidt and J.Soper hep-ph/9703411

[17] Particle Data Group, Phys.Lett.B **592**, 1 (2004).

[18] Jan Bartelski, Phys.Rev.D **71**, 014019 (2005).

[19] Hermes Collaboration, A.Airapetian, Phys.Rev.Lett **92**, 012005 (2004).

[20] R. Alkofer, H Reinhardt, and H Weigel, Phys.Rep **265**, 139 (1996).

[21] L. Gamberg, H. Weigel, and H: Reinhardt, Phys.Rev.D **58**, 054014 (1998).

[22] L. Gamberg, Structure Functions and Chiral Odd quark Distributions in the NJL, Riken BNL Research center , (2000).

[23] S. Aoki, et al., Phys.Rev.D **56**, 433 (1997).

[24] H.X.He " Quark Contributions to the Proton Spin and Tensor Charge" hep-ph/9712272

[25] H. -C.Kim, M Polyakov, and K. Goeke, Phys.Rev.D **53**, 4718 (1994); M. Wakamatsu and T. Kubota, Phys.Rev.D **60**, 034020 (1999).

[26] R.L. Jaffe and X. Ji, Nucl.Phys.B **375**, 527 (1992).

[27] H. He and X. Ji, Phys.Rev.D **52**, 2960 (1995); H. He and X. JI, Phys.Rev.D **56**, 6897 (1996).

[28] K. Suzuki and T. Shigetani, Nucl.Phys.B **626**, 886 (1997).

[29] Massimo Casu and L.M. Sehgal, Phys.Rev.D **55**, 2644 (1997).

[30] X.B. Chen, X.S. Chen, and F. Wang, Phys.Rev.Lett **87**, 012001 (2001); X.Chen et al., hep-ph/0003158